\documentclass[aps,prl,reprint,showpacs,showkeys,superscriptaddress,floatfix,twocolumn]{revtex4-2}

\usepackage{graphicx}
\usepackage{dcolumn}
\usepackage{bm}
\usepackage{hyperref}
\usepackage[mathlines]{lineno}
\usepackage{multirow}
\usepackage{natbib}
\usepackage{amsmath}
\usepackage{times}
\usepackage{xcolor}
\usepackage{soul}
\usepackage[normalem]{ulem}
\usepackage{textcomp,mathcomp}
\usepackage{bbold}
\usepackage{mathtools}
\usepackage{braket}
\usepackage{soul}

\newcommand{\YZGO}{YbZn$_2$GaO$_5$}
\newcommand{\LZGO}{LuZn$_2$GaO$_5$}

\begin{document}

\title{Evidence of Dirac Quantum Spin Liquid in YbZn$_2$GaO$_5$}

\author{Rabindranath Bag}
\affiliation{Department of Physics, Duke University, Durham, North Carolina 27708, USA}

\author{Sijie Xu}
\affiliation{Department of Physics, Duke University, Durham, North Carolina 27708, USA}

\author{Nicholas E. Sherman}
\affiliation{Department of Physics, University of California, Berkeley, California 94720, USA}
\affiliation{Materials Sciences Division, Lawrence Berkeley National Laboratory, Berkeley, California 94720, USA}

\author{Lalit Yadav}
\affiliation{Department of Physics, Duke University, Durham, North Carolina 27708, USA}

\author{Alexander I. Kolesnikov}
\affiliation{Neutron Scattering Division, Oak Ridge National Laboratory, Oak Ridge, Tennessee 37831, USA}

\author{Andrey A. Podlesnyak}
\affiliation{Neutron Scattering Division, Oak Ridge National Laboratory, Oak Ridge, Tennessee 37831, USA}

\author{Eun Sang Choi}
\affiliation{National High Magnetic Field Laboratory and Department of Physics, Florida State University, Tallahassee, Florida 32310, USA}

\author{Ivan da Silva}
\affiliation{ISIS Neutron and Muon Source, Rutherford Appleton Laboratory, Didcot, Oxfordshire OX11 0QX, UK}

\author{Joel E. Moore}
\affiliation{Department of Physics, University of California, Berkeley, California 94720, USA}
\affiliation{Materials Sciences Division, Lawrence Berkeley National Laboratory, Berkeley, California 94720, USA}

\author{Sara Haravifard}
\email[email:]{sara.haravifard@duke.edu} \affiliation{Department of Physics, Duke University, Durham, North Carolina 27708, USA} 
\affiliation{Department of Mechanical Engineering and Materials Science, Duke University, Durham, North Carolina 27708, USA}

\begin{abstract}
{The emergence of a quantum spin liquid (QSL), a state of matter that can result when electron spins are highly correlated but do not become ordered, has been the subject of a considerable body of research in condensed matter physics \cite{balentsNature2010, broholm2020quantum}.  Spin liquid states have been proposed as hosts for high-temperature superconductivity \cite{andersonScience1987} and can host topological properties with potential applications in quantum information science \cite{kitaevAnn2003}. The excitations of most quantum spin liquids are not conventional spin waves but rather quasiparticles known as spinons, whose existence is well established experimentally only in one-dimensional systems; the unambiguous experimental realization of QSL behavior in higher dimensions remains challenging.  Here we investigate the novel compound \YZGO, which hosts an ideal triangular lattice of effective spin-1/2 moments with no detectable inherent chemical disorder.  Thermodynamic and inelastic neutron scattering (INS) measurements performed on high-quality single crystal samples of \YZGO~exclude the possibility of long-range magnetic ordering down to 0.06 K, demonstrate a quadratic power law for the specific heat and reveal a continuum of magnetic excitations in parts of the Brillouin zone.  Both low-temperature thermodynamics and INS spectra suggest that \YZGO~is a U(1) Dirac QSL with spinon excitations concentrated at certain points in the Brillouin zone. We advanced these results by performing additional   specific heat measurements under finite fields, further confirming the theoretical expectations for a Dirac QSL on the triangular lattice of \YZGO.}
\end{abstract}
                              
\maketitle

\section{Introduction}\label{intro}
Anderson's proposal in 1973 ignited a surge of both experimental and theoretical efforts to identify the origins and properties of QSL states \cite{andersonMRB1973}. QSLs are exotic states of matter that remain disordered due to strong quantum fluctuations even at ultra-low temperatures \cite{balentsNature2010,savaryRPP2016,broholm2020quantum}. Despite significant experimental efforts, the unambiguous realization of a QSL state in the real world remains a challenge. In recent years, two-dimensional triangular lattice systems with rare-earth ions carrying effective spin-1/2 moments have presented promising opportunities in realizing QSL states, given the presence of spin-orbit coupling, crystal electric fields, and strong quantum fluctuations. Among these systems, the Yb-based YbMgGaO$_4$ has been intensively studied due to the absence of magnetic ordering and the observed continuum-like inelastic neutron scattering (INS) spectra, making it a promising candidate for QSL\cite{yueshengSR2015, yueshengPRL2015, yueshengPRL2016, shenNature2016, paddisonNP2017, ZhuPRL2017, KimchiPRX2018}. However, the presence of chemical disorder in YbMgGaO$_4$, caused by the inherently mixed occupancies of magnesium and gallium atoms on the same crystallographic site, has made the interpretation of the results challenging \cite{ZhuPRL2017, yueshengPRL2017, KimchiPRX2018}. Specifically, a theoretical study suggests that the chemical disorder may imitate the continuous INS spectra \cite{ZhuPRL2017}. Further studies on a sister compound, YbZnGaO$_4$, have proposed the presence of a spin-glass ground state attributed to the coexistence of chemical disorder and quantum fluctuation \cite{maPRL2018}. Therefore, eliminating or suppressing chemical disorder and accessing the intrinsic physics of an ideal triangular lattice of effective spin-1/2 moments is highly desired. As such, a potential candidate for hosting a QSL state is another class of Yb-based triangular lattice rocksalt-type compounds that do not have significant intrinsic chemical disorder: AYbX$_2$~ (A = Li, Na, K, Rb, Cs, and X = O, S, Se) \cite{Bordelon_Nature_2019, Ding_PRB2019, bordelon2020spin, Guo_PRM_2020, Ranjith_PRB_2019, DaiPRX2021, RanjithPRB2019, Zhang_PRB_2021, Zhang_PRB_2022, xing2019crystal, scheieArxiv2023}. Nevertheless, the task of obtaining high-quality single-crystal samples for this particular family has posed significant challenges. As a result, most of the reported results have been derived from powder samples or small single crystals, rendering the interpretation of data quite challenging. Furthermore, it is worth noting that in this compound family, the inter-layer Yb-Yb distance is relatively shorter, which introduces a more three-dimensional magnetic structure. Therefore, the key to unlocking the mystery of the QSL state in these systems hinges on creating and synthesizing single crystal samples for a new category of Yb-based triangular lattice compounds. These compounds should lack significant intrinsic chemical disorder and showcase a more two-dimensional magnetic structure.

\begin{figure*}[htbp]\centering\includegraphics[width= 13.5cm]{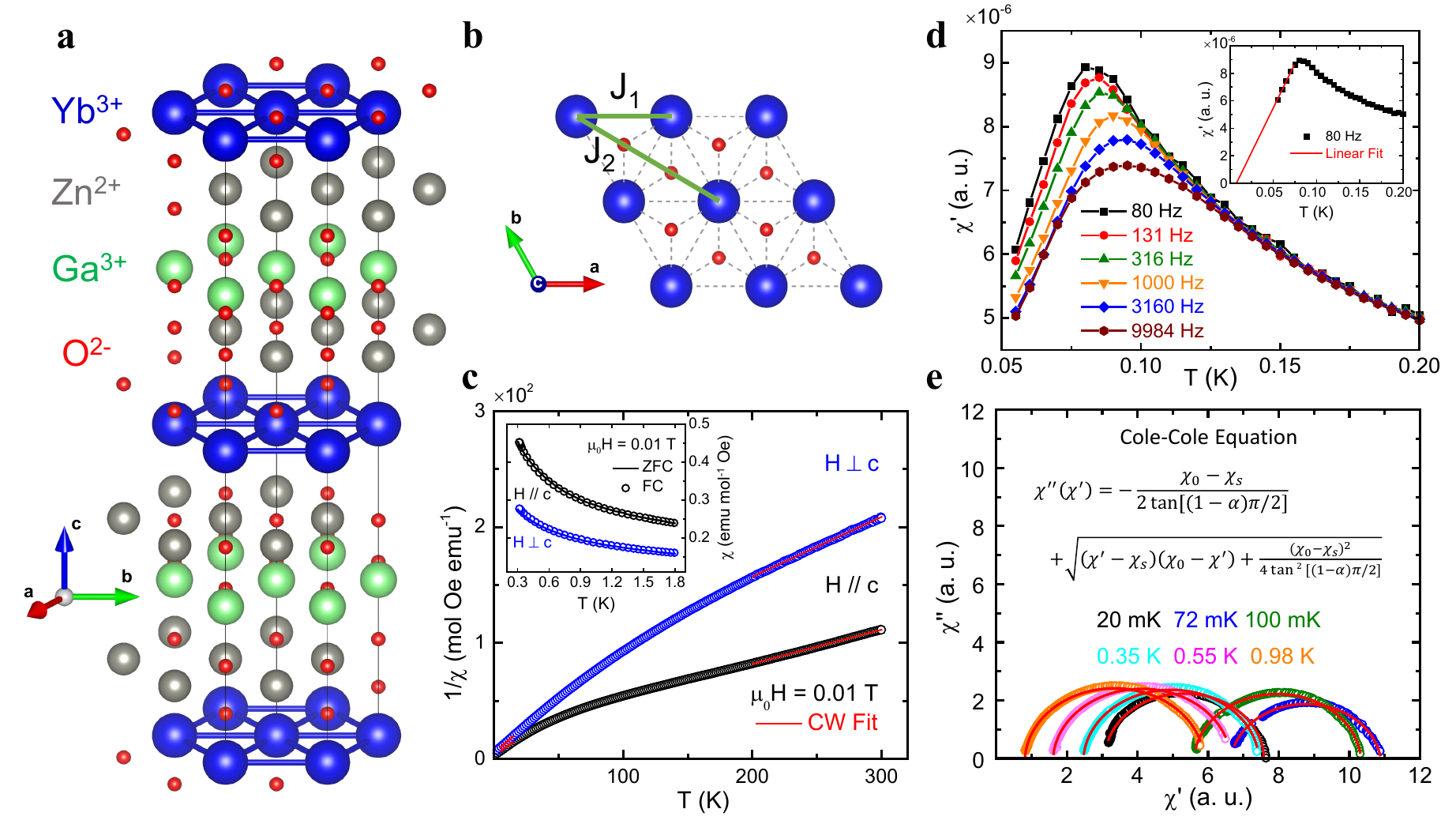} 

	\caption{\textbf{Crystal structure and magnetic susceptibility.} \textbf{a}. Crystal structure of \YZGO~; Yb-O planes are well separated by non-magnetic Zn-O, Ga-O, and Zn-O layers along crystallographic \textit{c}-direction. \textbf{b}. The Yb$^{3+}$ (blue sphere) forms a triangular lattice. The nearest neighbor couplings $J_1$ and next-nearest neighbor couplings $J_2$ are shown by green solid lines. \textbf{c}. The inverse magnetic susceptibility, $1/\chi$ ( H$\parallel$$c$ and H$\perp$$c$) data collected on single crystal sample of \YZGO~from 2 to 300 K. The red solid lines are the Curie-Weiss (CW) Fit at the low-temperature range from 5 - 15 K and at high-temperature range from 200 - 300 K. The inset shows no splitting between zero-field-cooling (ZFC) and field cooling (FC) magnetic susceptibility data of \YZGO~crystal down to 0.3 K. The measurements were conducted under an applied magnetic field of 0.01 T parallel and perpendicular to crystallographic $c$ - direction. \textbf{d}. Temperature dependence of the real part of the ac-susceptibility.  The inset displays a linear fit of the data at 80 Hz, ranging from 0.055 K to 0.08 K. Our linear fit extrapolates to the origin within the uncertainty range. \textbf{e}. Cole-Cole plot (Argand diagram) at different temperatures. The data were collected at a series of frequencies in the range of 10 to 3000 Hz. Data at each fixed temperature were individually fitted to the Cole-Cole equation shown in the plot. Our fit yields Cole-Cole coefficient $\alpha$ of 0.005(2), 0.080(1), 0.041(3), 0.033(1), 0.032(2), and 0.030(2), for 20 mK, 72 mK, 100 mK, 0.35 K, 0.55 K, and 0.98 K, respectively.}
	\label{Fig 1}
\end{figure*} 

To that end, we present a new Yb-based compound, \YZGO, that features an ideal triangular lattice of effective spin-1/2 moments without detectable intrinsic chemical disorder. Experiments aimed at elucidating the nature of the possible triangular lattice QSL in \YZGO~are of great importance. Thus, we grew a large and high-quality single crystal of \YZGO~(see Fig. S1) using the optical floating-zone technique to facilitate such experiments. Our specific heat measurements indicate that at ultra-low temperatures, the specific heat displays a $\sim$ T$^2$ dependence, indicating a U(1) Dirac QSL behavior \cite{RanPRL2017, HuPRL2019}. Additionally, we show that the field-induced T-linear component of the specific heat is proportional to the applied magnetic field, further confirming that the ground state of \YZGO~is best described with U(1) Dirac QSL \cite{RanPRL2017}.  In addition to the specific heat measurements, we conducted INS investigation on \YZGO~which reveals gapless, continuum-like spectra at the high-symmetry M and K points, but not throughout the Brillouin zone, and in particular not at the $\Gamma$ point. This particular pattern of low-energy spinon excitations is expected for the U(1) Dirac QSL phase \cite{ShermanPRB2023,drescher2022dynamical} and not for a spinon Fermi surface state.  Hence the specific heat scaling and INS spectra independently are best explained by low-energy spinon excitations with Dirac-like spectrum.

\section{Results \& Discussion}\label{results}
Before discussing the results further, we would like to review briefly the key signatures that differentiate a spinon Fermi surface state from a U(1) Dirac spin liquid. First, let us start with the spinon Fermi surface state. In such a state, the low temperature specific heat is predicted to scale as $\sim T$ \cite{YizhouRMP2017} at the mean-field level, possibly becoming $T^{2/3}$ if emergent gauge fluctuations dominate. As for the spectrum, mean-field theory predicts a V-shape spectrum near the $\Gamma$ point and gapless excitations throughout the entire Brillouin zone \cite{shenNature2016, LiPRB2017, savaryRPP2016}. These signatures have been observed in YbMgGaO$_4$ \cite{shenNature2016, ShenNC2018}, and NaYbSe$_2$ \cite{DaiPRX2021}. For the U(1) Dirac spin liquid, the specific heat is expected to scale as $\sim T^2$ because of the Dirac nodes \cite{RanPRL2017, PhysRevB.65.165113}. The low-energy theory on the triangular lattice is quantum electrodynamics ($\mathrm{QED}_3$) with 4 Dirac fermions ($N_f=4$), which predicts gapless excitations at both the $M$ and $K$ wavevectors  \cite{Song2019-2}. The main signature to distinguish this from a spinon Fermi surface state is the presence of a gap away from these points, such as near the $\Gamma$ point, as observed in recent simulations of the $J_1-J_2$ Heisenberg model on the triangular lattice \cite{Ferrari2019, ShermanPRB2023, drescher2022dynamical}.

Returning to the discussion of \YZGO, we demonstrate in Fig. \ref{Fig 1}a the hexagonal crystal structure of \YZGO, with the space group P6$_3$mmc. This is in contrast to YbMgGaO$_4$ and YbZnGaO$_4$ which crystallize in the R$\bar{3}$m space group. Within the P6$_3$mmc space group of \YZGO, distinctive Wyckoff positions are allocated for gallium and zinc. Consequently, \YZGO~exhibits no observable intrinsic chemical site mixing. The phase purity of our sample is confirmed through a Rietveld refinement analysis performed on the powder X-ray diffraction pattern obtained from a crushed \YZGO~single crystal (see Fig. S1). We further performed the single-crystal X-ray diffraction measurement on \YZGO~crystal and observed no chemical disorder within the limit of our experimental accuracy (see Table S1). In order to further confirm this, we also performed high-resolution neutron powder diffraction measurements on \YZGO. These results also show no signature of detectable site mixing in this system. This is an important observation since unlike their similar X-ray cross sections, the neutron cross sections for Zn and Ga differ by a factor of 1.65 (see Fig. S2), making it easier to detect the possible chemical site mixing. It is noteworthy to add that in \YZGO, an additional non-magnetic Zn-O layer is introduced along the crystallographic \textit{c}-direction, which increases the distance between magnetic Yb-O planes from 8.38$\AA$ (YbZnGaO$_4$) to 10.98$\AA$, enhancing the two-dimensionality and quantum fluctuations in this compound compared to previously reported Yb-based triangular lattice QSL candidates \cite{paddisonNP2017, maPRL2018, Bordelon_Nature_2019, Ding_PRB2019}. The nearest neighbor Yb$^{3+}$ ions are arranged in a perfect triangular pattern with a distance of 3.37$\AA$ between them, as illustrated in Fig. \ref{Fig 1}b. This distance is comparable to that in the previously reported Yb-based triangular lattice compounds proposed to host QSL state \cite{yueshengPRL2015, yueshengSR2015}.

\begin{figure}[htbp]\centering\includegraphics[width = 8 cm]{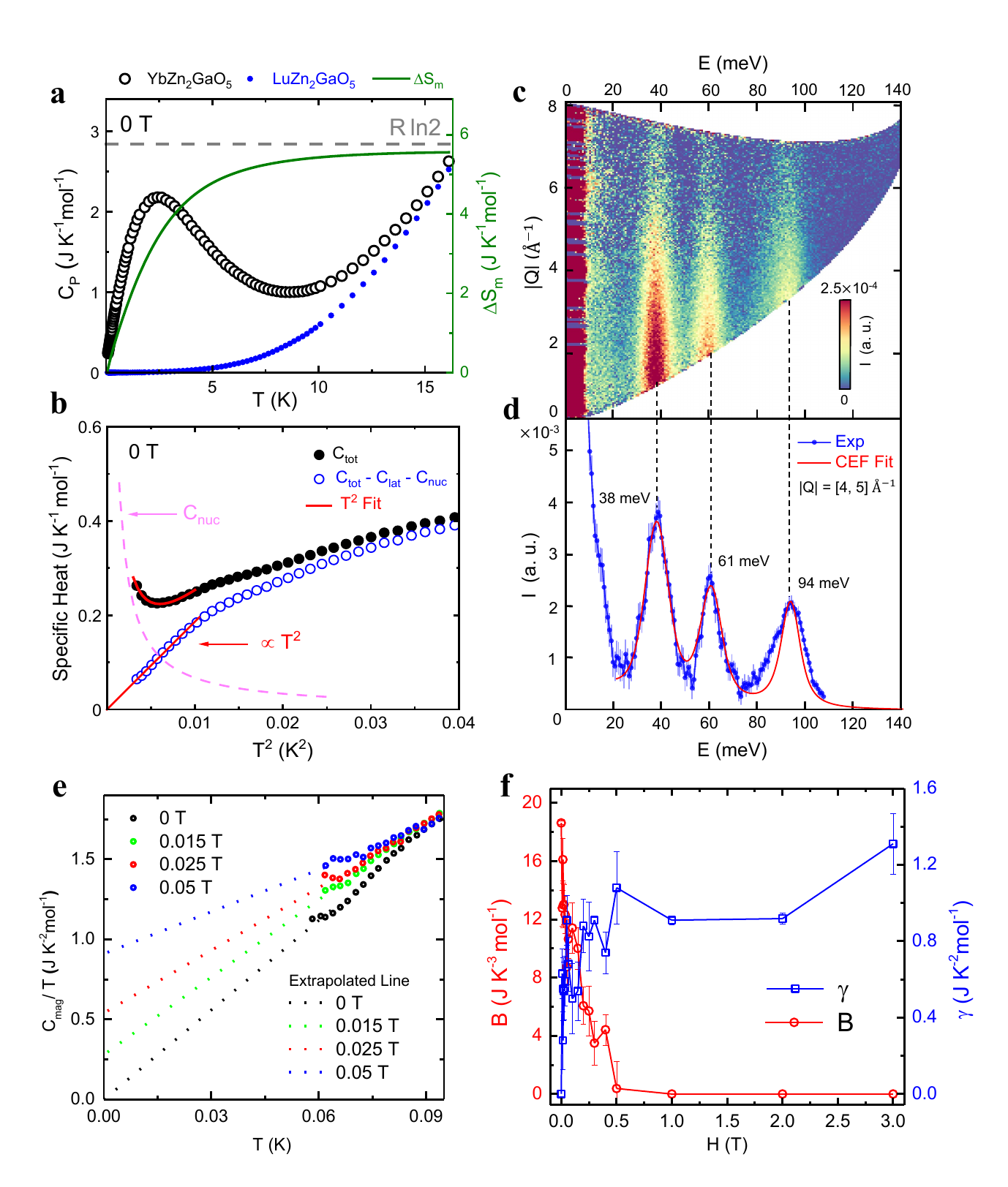}
	\caption{\textbf{Specific heat and crystal electric field levels.}  \textbf{a}. Specific heat data of \YZGO~single crystal and \LZGO~powder sample collected under zero field and down to 0.06 K are shown. The calculated magnetic entropy (right Y-axis) of \YZGO~saturates to Rln2, indicating the effective spin-1/2 ground state. \textbf{b}. Specific heat vs $T^2$ plot: Low-temperature total specific heat ($C_{tot} = C_{lat} + C_{nuc} + C_{mag}$) shows an upturn below 0.1 K. Using an isostructural nonmagnetic \LZGO~and a fitted Schottky model (C$_{nuc}$ $\simeq$ $AT^{-2}$), the lattice and nuclear contributions were removed. The magnetic specific heat data (C$_{mag}$) after subtraction is presented as open circles and fitted with a solid red straight line that exhibits a quadratic T-dependence ($C_{mag} \simeq B T^{2}$). The quadratic T-dependence fit yields an adjusted R-square value of 0.9583, indicating noticeable agreement between the model and experimental data. The T$^2$ dependence of magnetic specific heat data for T $\rightarrow$ 0 implies \YZGO~is a U(1) Dirac QSL candidate. \textbf{c} Inelastic neutron scattering (INS) spectra of \YZGO~reveal three crystal electric field (CEF) bands. The phononic contribution is subtracted using an isostructural nonmagnetic sample \LZGO.\textbf{d}. The single-ion CEF fitting shows energy levels at 38 meV, 61 meV, and 94 meV. The vertical error bars indicate statistical errors of one standard deviation. \textbf{e}. The magnetic specific heat (C$_{mag}$/T) under finite applied magnetic field plotted and extrapolated (dashed line) down to 0 K suggesting the Dirac QSL behavior, as  C$_{mag}$/T is linearly going up with magnetic field \cite{RanPRL2017}. \textbf{f}. The field dependence of $B$ and $\gamma$ fit parameters along with the error bars, extracted from the magnetic contribution of the specific heat modelled by $C_{mag} \simeq B T^{2} + \gamma T$.
	\label{Fig 2}}
\end{figure}
We illustrate in Fig. \ref{Fig 1}c the temperature-dependent magnetic susceptibility data collected on the single crystal of \YZGO~along both directions parallel and perpendicular to the crystallographic c-axis, in the presence of an applied magnetic field of 0.01 T. The inset of Fig. \ref{Fig 1}c highlights the low-temperature susceptibility region down to 0.3 K, confirming the absence of magnetic ordering down to this temperature. The inverse magnetic susceptibility data are fitted to Curie-Weiss law: $1/\chi$ = $(T - \theta_{CW})/C$, (where $\theta_{CW}$ is the Curie-Weiss temperature, and $C$ is the Curie constant) at two different temperature regimes. The obtained effective moment ($\mu_{eff}$ $\simeq$ 4.36 $\mu_B$) at a high-temperature range agrees with the expected theoretical value of free Yb$^{3+}$ ion ($\simeq$ 4.54 $\mu_B$). The Curie-Weiss temperatures obtained for \YZGO~in the low-temperature range (5 - 15 K) are slightly higher ($\theta_{CW,\parallel}$ = - 3.77 K and $\theta_{CW,\perp}$ = - 5.22 K) than those of Yb(Mg, Zn)GaO$_4$ \cite{yueshengSR2015, yueshengPRL2015, shenNature2016, XuPRL2016, maPRL2018}. This observation suggests a stronger antiferromagnetic coupling between Yb$^{3+}$ ions in \YZGO. The collected isothermal magnetization data along both directions up to 14 T of applied magnetic field provides anisotropic Landé-g factors of g$_\parallel$ $\simeq$ 3.44 and g$_\perp$ $\simeq$ 3.04 (see Fig. S1f).

As proposed in previous studies on Yb(Mg,Zn)GaO$_4$, one plausible explanation for the absence of long-range magnetic ordering could be that the ground state is a spin-glass \cite{maPRL2018}. In the spin-glass state, the spins are locked into a static, short-range ordered state below the freezing temperature (T$_f$) due to competing exchange interactions \cite{Mydosh1993}. Detection of the spin-glass state typically involves ac-susceptibility measurements. We conducted ac-susceptibility measurements on \YZGO~single crystal samples. In Figure \ref{Fig 1}d, the real part of the ac-susceptibility ($\chi'$) at 80 Hz reveals a maximum at $T_f \approx$ 0.082 K, at a slightly lower value compared to that observed in Yb(Mg, Zn)GaO$_4$ \cite{maPRL2018}. The observed peak exhibits a small frequency-dependent behavior, shifting slightly towards higher temperatures for low frequency regime. We attribute this behavior to the freezing of the orphan spins present in the small impurity phase rather than the spin-glass nature of the ground state for \YZGO. To further elaborate this, we show in the inset of  Fig. \ref{Fig 1}d the interpolation of lowest measured frequency (80 Hz) at the zero temperature limit, suggesting that the susceptibility passes through origin and is linear as a function of temperature in this regime. This is in agreement with our expectation for a Dirac quantum spin liquid state \cite{RanPRL2017, PhysRevB.65.165113} and implies that \YZGO~is unlikely to exhibit a spin-glass ground state. Furthermore, the spin-glass scenario is ruled out by the Cole-Cole analysis shown in Fig. \ref{Fig 1}e, where $\chi''$ versus $\chi'$ is plotted as frequency sweeps at fixed temperatures. The Cole-Cole plot illustrates the distribution of relaxation times, characterized by the fitting parameter $\alpha$. A perfect semi-circle corresponds to $\alpha = 0$ and implies a single relaxation time, while a highly flattened semi-circle corresponds to $\alpha \to 1$, indicating a broad distribution of relaxation times \cite{Cole1941}. In a typical spin-glass, one would anticipate a broad spectrum of relaxation times \cite{Mydosh1993, Binder1986}. However, in \YZGO, the obtained $\alpha$ falls within the range of 0.005 to 0.08, consistent with a single relaxation time at temperatures both below and above the peak position at $T_f$. Remarkably, even at the base temperature of 0.02 K, which is significantly below $T_f$, the smallest $\alpha$ value of 0.005 was obtained. These observations exclude the possibility of the spin-glass scenario and suggest that the ground state of \YZGO~remains dynamic down to 0.02 K. 

Theoretical models at the mean-field level predict Dirac QSL behavior at low temperatures, specifically when $k_BT \ll \chi J$, with $\chi$ representing the magnitude of the self-consistent mean-field parameter \cite{RanPRL2017}. According to this theory one expects specific heat $\propto T^2$ law due to the Dirac nodes. We show in Fig. \ref{Fig 2}a zero-field specific heat measurement performed on \YZGO~single crystal sample down to 0.06 K, confirming the absence of any long-range magnetic ordering. The contribution of phonons is subtracted using the isostructural non-magnetic sample of LuZn$_2$GaO$_5$. In Fig. \ref{Fig 2}b, we highlight the low-temperature behavior of the specific heat data, to comply with the $k_BT \ll \chi J$ condition. The specific heat of \YZGO~at this temperature range has multiple contributions denoted by $C_{tot} = C_{lat} + C_{nuc} + C_{mag}$. $C_{lat}$ represents the lattice (phononic) contribution that is negligible below 0.1 K. $C_{nuc}$ accounts for the nuclear Schottky contribution due to the hyperfine splitting of the nuclear energy levels resulting from the interaction between nuclear spins and electrons \cite{Gopal2012}. The nuclear contribution ($C_{nuc} \simeq AT^{-2}$) is subtracted by fitting the specific heat data below 0.1 K and the fitting coefficient value ($A$ = 6.72(4)$\times$ 10$^{-4}$ JKmol$^{-1}$) is consistent with that reported for other Yb-based compounds \cite{RanjithPRB2019}. Interestingly for \YZGO~ the subtracted magnetic specific heat data ($C_{mag}$) displays a $\propto T^2$ dependence (red solid straight line in Fig. \ref{Fig 2}b) with a coefficient of $\gamma = C_{mag}/T^2 = 18.6(1)$ JK$^{-3}$mol$^{-1}$ below 0.1 K. As discussed above, this is in agreement with the proposed Dirac QSL behavior \cite{RanPRL2017}. We also attempted to fit the magnetic specific heat data with a linear dependence ($C_{mag} \propto T$), but this fitting approach did not provide a satisfactory result (see Fig. S4). Therefore, the magnetic specific heat data exhibiting a $\propto T^2$ dependence for T $\rightarrow$ 0 and the lack of any long-range magnetic ordering down to the lowest temperature of 0.06 K imply that the ground state of \YZGO~is likely a U(1) Dirac QSL \cite{RanPRL2017, PhysRevB.65.165113}.

Another hallmark of Dirac spinons is their behavior in a Zeeman field, where the specific heat displays a linear increase with the applied field strength. In other words, at $k_BT \ll \mu_BH$, Dirac spinons are theorized to form a Fermi pocket where the magnetic field strength correlates directly with the pocket's radius. To explore this phenomena further, we measured the specific heat of \YZGO~ under various finite magnetic fields. To test this prediction, we gathered specific heat data at magnetic field strengths of 0.01 T, 0.015 T, 0.025 T, and 0.05 T, incrementing the temperature gradually. The results, illustrated in Fig. \ref{Fig 2}e, support the theoretical framework, showing a linear rise in (C$_{mag}$/T) as the magnetic field strength increases. This observed behavior stems from the splitting and vertical shift of Dirac nodes, which leads to a finite density of states (DOS) directly proportional to the field strength. Detailed specific heat results under different magnetic fields and a comparison between experimental fit parameters and theoretical predictions are presented in Fig. S5.

To provide further insight into the behavior of the specific heat as function of field, we model the magnetic contribution of the specific heat by $C_{mag} \simeq B T^{2} + \gamma T$. We show in Fig. \ref{Fig 2}f the field dependence of the extracted $B$ and $\gamma$ parameters. The results obtained for $B$ confirm the $T^{2}$ behavior of the specific heat at zero field, while the values extracted for $\gamma$ demonstrate a linear increase in specific heat as a function of magnetic field. The field-dependent ratio of $\gamma$ and $B$ coefficients is also consistent with theoretical expectations (see Supplementary Material), further confirming that the ground state of \YZGO~is best described by U(1) Dirac QSL.

\begin{figure*}[htbp]\centering\includegraphics[width= 13.5 cm]{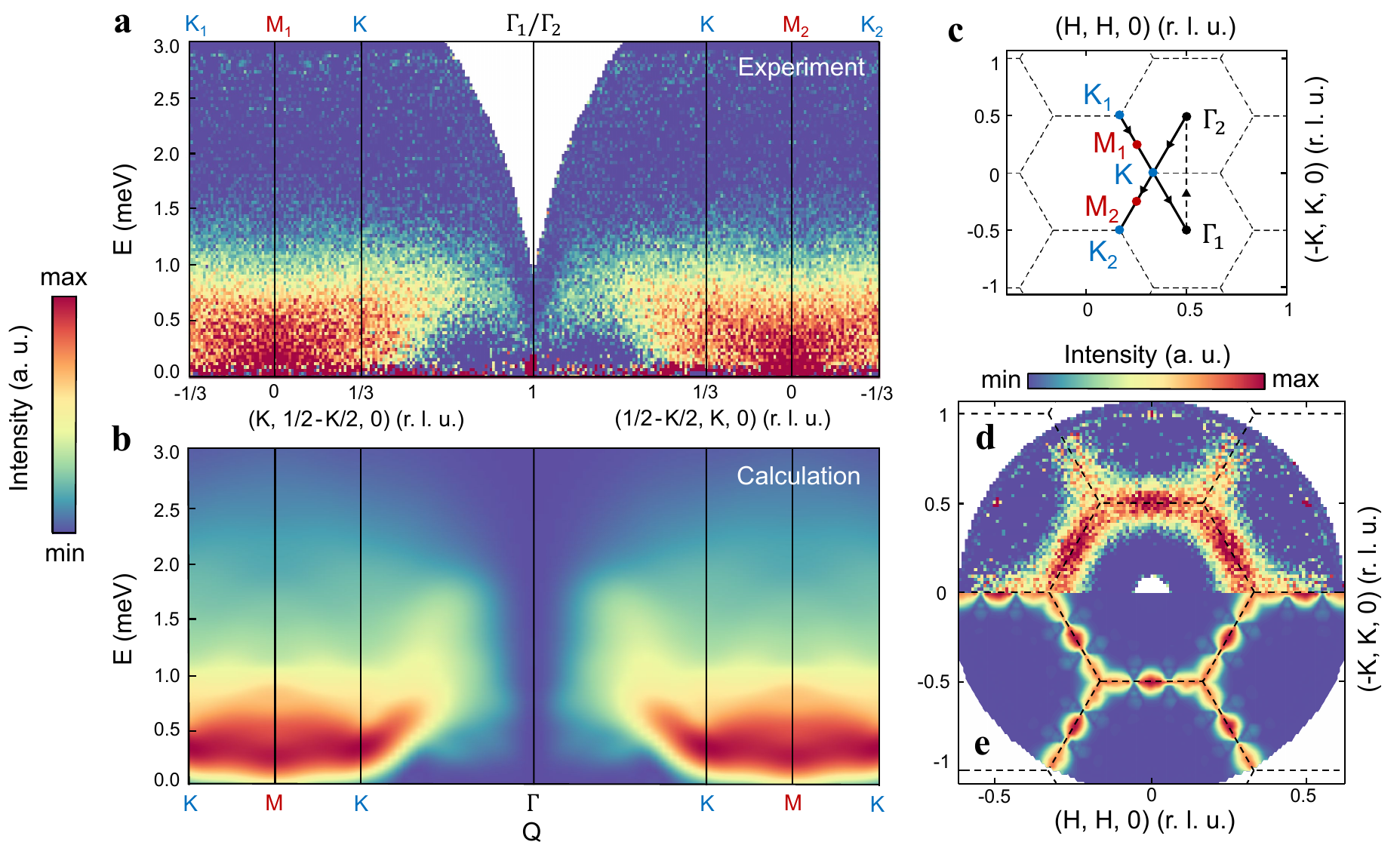}  
	\caption{\textbf{Inelastic Neutron scattering data under zero applied magnetic field.} \textbf{a}. Energy dependence of the magnetic excitation spectrum along high symmetry points measured at 0.1 K. The background is subtracted using the high temperature (45 K) spectrum (See Methods section for details). The contour path travels along the high symmetry points $K_1$-$M_1$-$K$-$\Gamma_1/\Gamma_2$-$K$-$M_2$-$K_2$, which is illustrated by the black solid curve in c. \textbf{b}. Calculated spectrum using matrix product states for the $J_1-J_2$ XXZ model on the triangular lattice, with $J_2 / J_1 = 0.12$ and $\Delta = 1.35$ (see Methods section for details). We use J$_1$ $\sim$ 0.5 meV to adjust the scale of the y-axis for better comparison of the experimental data. \textbf{c}. A schematic of the high symmetry path used in a,b. The dashed lines show the boundary of the Brillouin zones. \textbf{d}. Background-subtracted low energy slice of the magnetic excitation spectrum collected at 0.1 K. The energy integration range is [0.1, 0.3] meV. \textbf{e}. The calculated spectrum of the $J_1-J_2$ XXZ model with the same parameters as in b. We integrate the spectrum in the energy range [0.1, 0.3] meV for comparison with d.
}\label{Fig 3}
\end{figure*} 

To gain deeper insights into the dynamics of \YZGO, we conducted an inelastic neutron scattering (INS) experiment using a high-purity powder sample at 5 K, as depicted in Fig. \ref{Fig 2}c,d. The single-ion crystal electric field (CEF) fitting results revealed three distinct crystal field excitations at 38 meV, 61 meV, and 94 meV, which agree with those observed in other previously reported Yb-based triangular lattice systems \cite{yueshengPRL2017, Ding_PRB2019, SarkarPRB2019, DaiPRX2021, ZhangPRB2021, SteinhardtnpjQM2021}. Based on our analysis of the CEF levels, we have determined that the first excited state of \YZGO~is separated from the ground state by a gap of over 441 K (38 meV). This suggests a Kramer's doublet ground state for the Yb$^{3+}$ ions with an effective spin-1/2. Our CEF fitting scheme, which is consistent with the specific heat data, supports this conclusion. In particular, the calculated magnetic entropy of \YZGO~saturates at Rln2, indicating an effective spin-1/2 ground state, as shown in Fig. \ref{Fig 2}a (see Tables S2 and S3 for further details on the CEF fitting scheme). 

It is important to note the observed broadening of the 94 meV CEF excitation band. Previous studies have linked broadening at 97 meV and the emergence of an additional band at 138 meV in YbMgGaO$_4$ to disorder \cite{yueshengPRL2017}. However, broadening in CEF levels can also stem from various mechanisms in rare-earth-based compounds. For instance, Gaudet \textit{et al.} \cite{GaudetPRB2018} describe a shoulder-like feature in one of the CEF levels of Ho$_2$Ti$_2$O$_7$, which results from the hybridization of a nearby phonon mode with the CEF mode, forming a vibronic mode. Consequently, the shoulder observed at approximately 87.5 meV in our data does not necessarily suggest disorder. In \YZGO, phonon excitations up to approximately 90 meV were also observed in the INS data collected on the isostructural non-magnetic compound \LZGO, which could potentially lead to the formation of vibronic bonds. Thus, we propose that the observed broadening may be attributed to CEF-phonon coupling, underscoring the need for further detailed investigation of this system using complementary techniques such as Raman spectroscopy \cite{Zhang_PRB_2021, PaiJMCC2022}.

Furthermore, we conducted INS experiments on a high-quality single crystal of \YZGO. These experiments were designed to probe the low-energy excitations of the material. An incident neutron energy of 3.32 meV was used and the excitation spectrum under zero-field was collected at a base temperature of 0.1 K. Scans performed at a high temperature of 45 K were used for background subtraction (see Fig. S3). We show in Fig.\ref{Fig 3}a the scattering intensity as a function of energy transfer, with a path taken through the high-symmetry points of $K_1$-$M_1$-$K$-$\Gamma_1$/$\Gamma_2$-$K$-$M_2$-$K_2$, as illustrated in Fig.\ref{Fig 3}c. 

Additionally, a broad continuum extending over an energy scale of approximately 1.4 meV was observed, with the spectrum weight gradually decreasing at higher energy. We observed a clear gap in the spectra of \YZGO~near the $\Gamma$ point, while the excitation appears to be remained gapless between M and K points within the instrumental energy resolution of approximately 0.06 meV, as shown in Fig. \ref{Fig 3}a. We present in Fig. \ref{Fig 3}b the outcome of our theoretical calculation, which are based on the J$_1$-J$_2$ XXZ model and use large-scale matrix product state simulations (see the Methods section for details) \cite{ShermanPRB2023}. These results demonstrate promising agreement with our experimental data, with J$_2$/J$_1$ $\sim$ 0.12 and an anisotropy value of $\Delta$ $\sim$ 1.35 (see Fig. S6 b). 

Even though the energy of the neutrons is higher than that needed to access the Dirac behavior directly, they do observe a spectrum that is consistent with the $J_1-J_2$ model in the regime in which the Dirac QSL is most likely the ground state. It is worth noting that on the triangular lattice, the energy scale of the spinons in $J_1-J_2$ model is not the largest magnetic energy scale $J_1$, which is the nearest-neighbor interaction, because on the triangular lattice, the $J_1$ coupling alone would lead to long-ranged three-sublattice order. The second largest coupling seems to be a second-neighbor $J_2$, based on linear-spin-wave fits at high magnetic fields for similar Yb-based triangular lattice compounds \cite{ShermanPRB2023, DaiPRX2021}, and the spinon energy scale is then expected to depend on the dimensionless parameter $J_2/J_1$. From numerical calculations, there is a window of $J_2/J_1$ between long-ranged ordered phases and where the QSL state is predicted to emerge; reasonable values are $J_2/J_1$ between 0.06 and 0.12. Therefore, the energy scale of the zone-boundary spinon is expected to be a lot less than $J_1$, which is consistent with a significant contribution to the specific heat at low temperatures and also with the violations of $T^2$ behavior at above $\sim$ 0.1 Kelvin.

In contrast to \YZGO, other reported Yb-based triangular systems, such as the delafossite material NaYbSe$_2$, display a continuum spectrum without a gap at all Q. This observation is consistent with a spinon Fermi surface state \cite{DaiPRX2021, LiPRB2017}. However, in \YZGO, the observed energy gap near $\Gamma$ along with the continuum spectra observed at Q = K and M are indicative of a U(1) Dirac QSL state \cite{ShermanPRB2023}. The thermodynamic measurements are consistent with this picture, as the magnetic specific heat data ($C_{mag}$) exhibits a $\propto T^2$ dependence for $T$ $\rightarrow$ 0 as well as $\propto T$ behavior as function of magnetic field, which are also indicative of a U(1) Dirac QSL.

In Fig. \ref{Fig 3}d, we demonstrate the experimental dispersion of the neutron scattering intensities through constant energy slice integrated from 0.1 to 0.3 meV. We observe that at low energy, spectral weights are localized at the edge of the Brillouin zone, with higher intensities at the $M$ points than at the $K$ points (see Fig. S6 a for linear cut). As the energy increases, the difference in intensities between $M$ and $K$ points becomes less prominent, and the intensities disperse throughout the Brillouin zone, as shown in Fig. S7. Moreover, the theoretical calculations in Fig. \ref{Fig 3}e agree with our experimental observations in similar energy ranges. The agreement between theory and experiment suggests that INS spectra of \YZGO~are sufficiently well described by the $J_1-J_2$ XXZ model without additional couplings. The presence of what appears to be gapless modes at $K$ and $M$ while having a gap near $\Gamma$ strongly favors a U(1) Dirac QSL ground state over either a spinon Fermi surface or the effects of disorder. 

\section{Conclusions}\label{conclusions}
In conclusion, we have successfully synthesized and characterized high-quality single crystals of \YZGO, a Yb-based triangular lattice system that displays no observable intrinsic chemical disorder. Our ac-susceptibility measurements rule out the possibility of a spin-glass ground state. Furthermore, our specific heat measurements reveal a quadratic temperature dependence of the magnetic component at ultra-low temperatures, as well as linear-$T$ dependence as function of magnetic field, providing strong experimental evidence for the presence of a U(1) Dirac QSL state on a triangular lattice in \YZGO. Additionally, our INS investigation of \YZGO~shows gapless, continuum-like spectra at the high-symmetry $M$ and $K$ points, but \textit{not} at the $\Gamma$ point, confirming the potential existence of a U(1) Dirac QSL ground state in this material, and these experimental findings align well with theoretical interpretations.  Consequently, we firmly believe that \YZGO~represents a highly promising candidate for the elusive U(1) Dirac QSL, and we advocate for further exploration using a diverse range of theoretical and experimental techniques. Furthermore, we have recently uncovered a substantial correlation between the predicted Dirac spin liquid spectrum for the superconducting cuprate pseudogap \cite{christos2023model} and the physical properties observed in our compound. This discovery shines a spotlight on the potential realization of a Dirac QSL in \YZGO, underscoring its significance within the field.

\section{Acknowledgements}\label{acknowledgements}
We would like to express our gratitude to Patrick A. Lee, Cristian D. Batista, and Sachith E. Dissanayake for their valuable contributions to our discussions. We are thankful to Chun-Hsing Chen and the X-ray Core Laboratory of the Department of Chemistry at the University of North Carolina at Chapel Hill for their help with single-crystal X-ray diffraction measurements. Work performed at Duke University is supported by the U.S. Department of Energy, Office of Science, Office of Basic Energy Sciences, under Award Number DE-SC0023405. Work performed at the University of California, Berkeley, is supported by the U.S. Department of Energy, Office of Science, National Quantum Information Science Research Centers, Quantum Science Center.  N.S. is additionally supported by the Theory Institute for Materials and Energy Spectroscopies (TIMES) of the U.S. Department of Energy, Office of Basic Energy Sciences, under contract DE-AC02-05CH11231.  R.B. acknowledges the support provided by Fritz London Endowed Post-doctoral Research Fellowship. This research used resources of the Spallation Neutron Source at Oak Ridge National Laboratory, which is a DOE Office of Science User Facility. A portion of this work was performed at the National High Magnetic Field Laboratory, which is supported by National Science Foundation Cooperative Agreement No. DMR-2128556 and the State of Florida. 

\section{Author Contributions}
Research conceived by S.H.; Samples synthesized by S.X., R.B., and S.H.; Magnetic measurements performed and analyzed by S.X., R.B., and S.H.; specific heat measurements performed and analyzed by S.X., R.B, and S.H.; ac-susceptibility measurements performed and analyzed by S.X., E.S.C., and S.H.; Neutron scattering measurements performed and analyzed by S.X., A.I.K., A.A.P, I.S., and S.H.; Theoretical calculations performed by N.E.S. and J.E.M.; Manuscript written by S.X., R.B., N.E.S., J.E.M., and S.H.; All authors commented on the manuscript.

\section{Data Availability}\label{data availability}
The data that support plots and other findings in this paper are available from the corresponding author upon reasonable request.

\section{Code Availability}\label{code availability}
The computer codes used to generate results are available from the corresponding author upon reasonable request.

\section{Appendixes}\label{methods}

\subsection{Appendix A: Sample preparation and single crystal growth}\label{sample prep}
The poly-crystalline sample of \YZGO~was synthesized using a solid-state reaction route. The high-purity precursors of Yb$_2$O$_3$ (99.9$\%$), Ga$_2$O$_3$ (99.9$\%$), and ZnO (99.9$\%$) with 5$\%$ excess ZnO were used and mixed thoroughly and then pressed into a pellet. The pellets were sintered at 1275 $\tccentigrade$ for 36 hours with intermediate grinding to obtain a pure phase of \YZGO. The phase purity is confirmed using Powder X-ray diffraction (PXRD) (see Fig. S1). The pure powder sample and around 10$\%$ excess of ZnO were mixed and pressed into a cylindrical rod using a hydrostatic pressure of 700 bar. Single crystals of \YZGO~were grown using the optical floating-zone technique in the presence of a 10-bar oxygen atmosphere. A transparent single-grain crystal was successfully obtained, with a cleaved facet plane along [001] which was confirmed from Laue X-ray diffraction (see Fig. S1).

\subsection{Appendix B: Single crystal X-ray diffraction}\label{scxrd}
The single crystal X-ray diffraction (SCXRD) is performed on \YZGO~single crystal sample at the Department of Chemistry, University of North Carolina. A colorless transparent crystal piece (approximate dimensions $0.020 \times 0.010 \times 0.010 ~mm^3 $) was harvested by an X-ray transparent loop made by MiTeGen and mounted on a Bruker D8 VENTURE diffractometer and measured at 150 K. The data collection was carried out using Mo-K$\alpha$ radiation (graphite monochromator) with a frame time of 0.4 seconds and a detector distance of 4 cm. 

\subsection{Appendix C: AC susceptibility} \label{AC}
The ac susceptibility experiments were carried out using two setups: The ac susceptibility $\chi'$ is measured in a dilution refrigerator attached to a Quantum Design (QD) Physical Property Measurement System (PPMS) Dynacool and the Cole-Cole plot (Argand diagram) is measured in the SCM1 at the National High Magnetic Field Laboratory (MagLab). For the first experiment carried out in a PPMS, a piece of high-quality single crystal with a mass of 14 mg was used. The temperature-dependent $\chi'$ measurements were conducted from 1 K to 0.055 K.  For the MagLab measurements, four pieces of high-quality single crystal samples, with a total mass of 25 mg, were co-aligned and stacked together to enhance the signal. The samples were then placed inside a Kapton tube and secured with a small amount of super glue. The single crystals were carefully polished with sandpaper to achieve a close-to-one filling factor within the tube which can help achieve the optimal resolution. For the Cole-Cole analysis, a sequence of frequencies from 40 Hz to 3000 Hz was used at different fixed temperatures.
   
\subsection{Appendix D: Specific heat measurements} \label{HC}
Specific heat measurements were carried out on  single crystal samples of \YZGO~and  LuZn$_2$GaO$_5$ using Helium-4 (1.8 K $\leq$ T $\leq$ 300 K) and dilution refrigerator (0.05 K $\leq$ T $\leq$ 2 K) set up attached to Quantum Design PPMS Dynacool. A representative single-crystal sample of \YZGO~mounted on a specific heat measurement platform is shown in the supplementary information (see Fig. S4).  

\subsection{Appendix E: Magnetic measurements}\label{magnetic}
Temperature-dependent magnetic susceptibility was measured using a 7 Tesla Cryogenic Ltd SQUID (superconducting quantum interference device) magnetometer with a Helium-3 probe from 0.3 K to 2 K and with a Helium-4 probe from 2 K to 300 K. For the Helium-3 measurements, a small crystalline \YZGO~sample of 1.04 mg was mounted on a silver sample holder and the Helium-4 measurement, 11.90 mg of \YZGO~single crystal was used. The crystals were oriented using the Laue diffraction method and the regular shape of the single crystals was obtained using a wire saw. The magnetic measurements were performed under an applied magnetic field parallel ($H$$\parallel$c) and perpendicular ($H$$\perp$c) to the crystallographic \textit{c}-directions of \YZGO. The isothermal magnetization measurements along both directions of \YZGO~single crystal sample were performed using a VSM (vibration sample magnetometer) in PPMS up to 14 Tesla of the applied magnetic field (see Fig. S1).   

\subsection{Appendix F: Neutron powder diffraction}\label{NPD}
Neutron Powder Diffraction (NPD) measurement of \YZGO~was collected using the General Materials Diffractometer (GEM) at the ISIS Neutron and Muon Source (Rutherford Appleton Laboratory, United Kingdom) \cite{HANNON200588}. The sample was loaded into an 8 mm diameter vanadium sample holder, and the NPD pattern was collected at room temperature, with a 15/40 mm (horizontal/vertical) beam size.

\subsection{Appendix G: Inelastic neutron scattering}\label{INS}
The inelastic neutron scattering (INS) experiments were performed on the Fine-Resolution Fermi Chopper Spectrometer (SEQUOIA) \cite{granroth2010sequoia} and the Cold Neutron Chopper Spectrometer (CNCS) \cite{ehlers2011cncs} at the Spallation Neutron Source (SNS), Oak Ridge National Laboratory. For the SEQUOIA experiment, 6.5 g of pure powder samples of \YZGO~and \LZGO~were used with incident neutron energies of $E_i=$ 80 and 120 meV at temperatures of $T=$ 5 and 100 K. The phonon contributions in the \YZGO~spectrum were subtracted using isostructural non-magnetic \LZGO. For the CNCS experiment, 10 pieces of high-quality single crystal samples of \YZGO~with a total mass of $\sim$ 1.8 g were co-aligned within 1.5$^{\circ}$ using a Laue X-ray back-scattering diffractometer and mounted along $(hk0)$ scattering plane on an oxygen-free copper sample holder (see Fig. S1). The measurements were carried out in a dilution refrigerator with a base temperature of 0.1 K. A neutron-absorbing Cd foil was placed at the bottom of the holder to reduce the background from the sample holder. The measurements were conducted at the base temperature and 45 K under zero-field with an incident neutron energy of $E_i=$3.32 meV. The sample was rotated with an increment of 1$^{\circ}$, with a range of -180$^{\circ}$ to 180$^{\circ}$. The data were analyzed using the HORACE software and were folded 3 times along the high symmetry axis [0$H$0], [$\bar{H}$$H$0], and [$H$00] into a 60 $^{\circ}$ sector in the reciprocal space to improve statistics. For the constant energy slice, the folded data were cut, duplicated, and recombined to restore 360$^{\circ}$ coverage for the purpose of presentation. To better extract the magnetic signal of interest, a comparative analysis was conducted between the 0.1 K and 45 K spectra. The 45 K spectrum was normalized to 0.1 K data and used as background which then was subtracted from the 0.1 K spectrum.

\subsection{Appendix H: Theoretical calculations}\label{sec:theory-methods}

We performed dynamical tensor network simulations of the $J_1-J_2$ XXZ model on the triangular lattice, given by
\begin{align}
    H = &J_1\sum_{\langle{i,j}\rangle}
    \left(\Delta S_i^zS_j^z +  S_i^xS_j^x +  S_i^yS_j^y\right)\\
    +&J_2\sum_{\langle\hspace{-2pt}\langle{i,j}\rangle\hspace{-2pt}\rangle}
    \left(\Delta S_i^zS_j^z +  S_i^xS_j^x +  S_i^yS_j^y\right),
    \label{eq:H_J1-J2}
\end{align}
where $S_i^{\alpha}$ are spin-$1/2$ operators, and $\langle{i,j}\rangle$  and $\langle\hspace{-2pt}\langle{i,j}\rangle\hspace{-2pt}\rangle$ denote nearest- and next-nearest neighbor pairs, respectively. In this work, we only look at the value of $J_2 / J_1 = 0.12$, which was chosen based on the similarity between the experimental spectra and the calculated spectra at this value reported in \cite{ShermanPRB2023}.  This value is believed to be deep in the QSL phase of the isotropic model \cite{Iqbal2016}.  To reflect some of the lowered symmetry in \YZGO, we add one anisotropy parameter $\Delta$ and study the $\Delta$ dependence; in principle, $J_1$ and $J_2$ could have different anisotropy parameters, and one could incorporate higher-order couplings as well, but the computational effort required to map the Brillouin zone for each choice of parameters means that we leave such further explorations for future work.

 The quantity of interest is the dynamical structure factor, given by
\begin{equation} \label{eq:Sqw}
    S\bigl(\boldsymbol{q},\omega\bigr) = 
    \frac{1}{N}\sum_{\boldsymbol{x}}\int_{0}^{\infty}\frac{\mathrm{d}t}{2\pi}e^{i\left(\omega t - \boldsymbol{q}\cdot \boldsymbol{x}\right)}G\bigl(\boldsymbol{x},t\bigr),
\end{equation}
with $N$ the number of lattice sites. The quantity $G\bigl(\boldsymbol{x},t\bigr)$ is the two-point spin-spin correlation function defined as
\begin{equation}
    G(\boldsymbol{x},t) = \bra{\Omega} \boldsymbol{S}_x(t) \cdot \boldsymbol{S}_c(0)\ket{\Omega},
\end{equation}
where $c$ denotes the center site of the lattice taken to be the origin, $\boldsymbol{x}$ is the distance of site $\boldsymbol{x}$ from the origin, and $\ket{\Omega}$ is the ground state of $H$ with energy $E_0$. To calculate this quantity, we split the calculation into three parts by writing
\begin{equation}
     \boldsymbol{S}_x \cdot \boldsymbol{S}_c = S_x^zS^z_c + \frac{1}{2}\left[S^+_xS^-_c + S^-_xS^+_c\right]
\end{equation}

To calculate $S\bigl(\boldsymbol{q},\omega\bigr)$, we use the same method described in Ref. \cite{ShermanPRB2023}, but we will provide a summary of the technique here. 

First, we curl the triangular lattice into a cylinder with a circumference $C=6$, and a length $L=36$, such that $N=LC$, using the XC boundary conditions \cite{Szasz2020}. Then, working with matrix product states (MPS),  we calculate the ground state using the density matrix renormalization group (DMRG) \cite{White1992, Schollwock2011}. The time evolution for $G(\boldsymbol{x},t)$ is calculated using the time-dependent variational principle (TDVP) \cite{PhysRevLett.107.070601, PhysRevB.88.075133, PhysRevB.94.165116, Vanderstraeten2019, Yang2020}, with a time step of $0.1$, and maximum time $T_\mathrm{max}=30$. We use a maximum bond dimension $\chi=512$ for the simulations. 

To smooth the data in frequency space, we rescale $G(\boldsymbol{x},t)$ by a Gaussian. In particular
\begin{equation}
    G(\boldsymbol{x},t) \longrightarrow  e^{-\eta t^2} G(\boldsymbol{x},t),
\end{equation}
and we use $\eta=0.02$ here. Since our system is inversion symmetric, we rewrite the Fourier transform in Eq. \eqref{eq:Sqw} as 
\begin{align}\label{eq:ft_final}
    S\bigl(\boldsymbol{q},\omega\bigr) &= \frac{1}{\pi N}\int_0^{\infty}\mathrm{d}t\sum\nolimits_{\boldsymbol{x}} \cos\bigl(\boldsymbol{q}\cdot\boldsymbol{x}\bigr)  \nonumber\\ &\times\Bigl(\cos(\omega t)\mathsf{Re} G(\boldsymbol{x},t) - \sin(\omega t)\mathsf{Im} G(\boldsymbol{x},t)\Bigr),
\end{align}
yielding the results displayed here. 

Lastly, due to a small circumference of the cylinder used during the simulations, this restricts the allowed wavevectors $\textbf{q}$ to a subset of the full Brillouin zone. Since the infinite system has a full $C_6$ rotational symmetry, this means that the dynamical structure factor should have the same symmetry. In particular, in the thermodynamic limit, for any rotation $R\in C_6$, we have
\begin{equation}
    S\bigl(R\boldsymbol{q},\omega\bigr) = S\bigl(\boldsymbol{q},\omega\bigr).
\end{equation}
We use this factor to generate the values $S\bigl(R\boldsymbol{q},\omega\bigr)$ from the initial data $S\bigl(\boldsymbol{q},\omega\bigr)$. For more details on the simulations, we point the reader to Ref. \cite{ShermanPRB2023}.

\bibliographystyle{apsrev4-2}
\bibliography{Yb1215_Main.bib}

\end{document}


\title{Supplementary materials for `Evidence of Dirac Quantum Spin Liquid in YbZn$_2$GaO$_5$'}

\author{Rabindranath Bag}
\affiliation{Department of Physics, Duke University, Durham, North Carolina 27708, USA}

\author{Sijie Xu}
\affiliation{Department of Physics, Duke University, Durham, North Carolina 27708, USA}

\author{Nicholas E. Sherman}
\affiliation{Department of Physics, University of California, Berkeley, California 94720, USA}
\affiliation{Materials Sciences Division, Lawrence Berkeley National Laboratory, Berkeley, California 94720, USA}

\author{Lalit Yadav}
\affiliation{Department of Physics, Duke University, Durham, North Carolina 27708, USA}

\author{Alexander I. Kolesnikov}
\affiliation{Neutron Scattering Division, Oak Ridge National Laboratory, Oak Ridge, Tennessee 37831, USA}

\author{Andrey A. Podlesnyak}
\affiliation{Neutron Scattering Division, Oak Ridge National Laboratory, Oak Ridge, Tennessee 37831, USA}

\author{Eun Sang Choi}
\affiliation{National High Magnetic Field Laboratory and Department of Physics, Florida State University, Tallahassee, Florida 32310, USA}

\author{Ivan da Silva}
\affiliation{ISIS Neutron and Muon Source, Rutherford Appleton Laboratory, Didcot, Oxfordshire OX11 0QX, UK}

\author{Joel E. Moore}
\affiliation{Department of Physics, University of California, Berkeley, California 94720, USA}
\affiliation{Materials Sciences Division, Lawrence Berkeley National Laboratory, Berkeley, California 94720, USA}

\author{Sara Haravifard}
\email[email:]{sara.haravifard@duke.edu} \affiliation{Department of Physics, Duke University, Durham, North Carolina 27708, USA} 
\affiliation{Department of Mechanical Engineering and Materials Science, Duke University, Durham, North Carolina 27708, USA}

\maketitle
	\begin{figure*}[htbp]\centering\includegraphics[width=16cm]{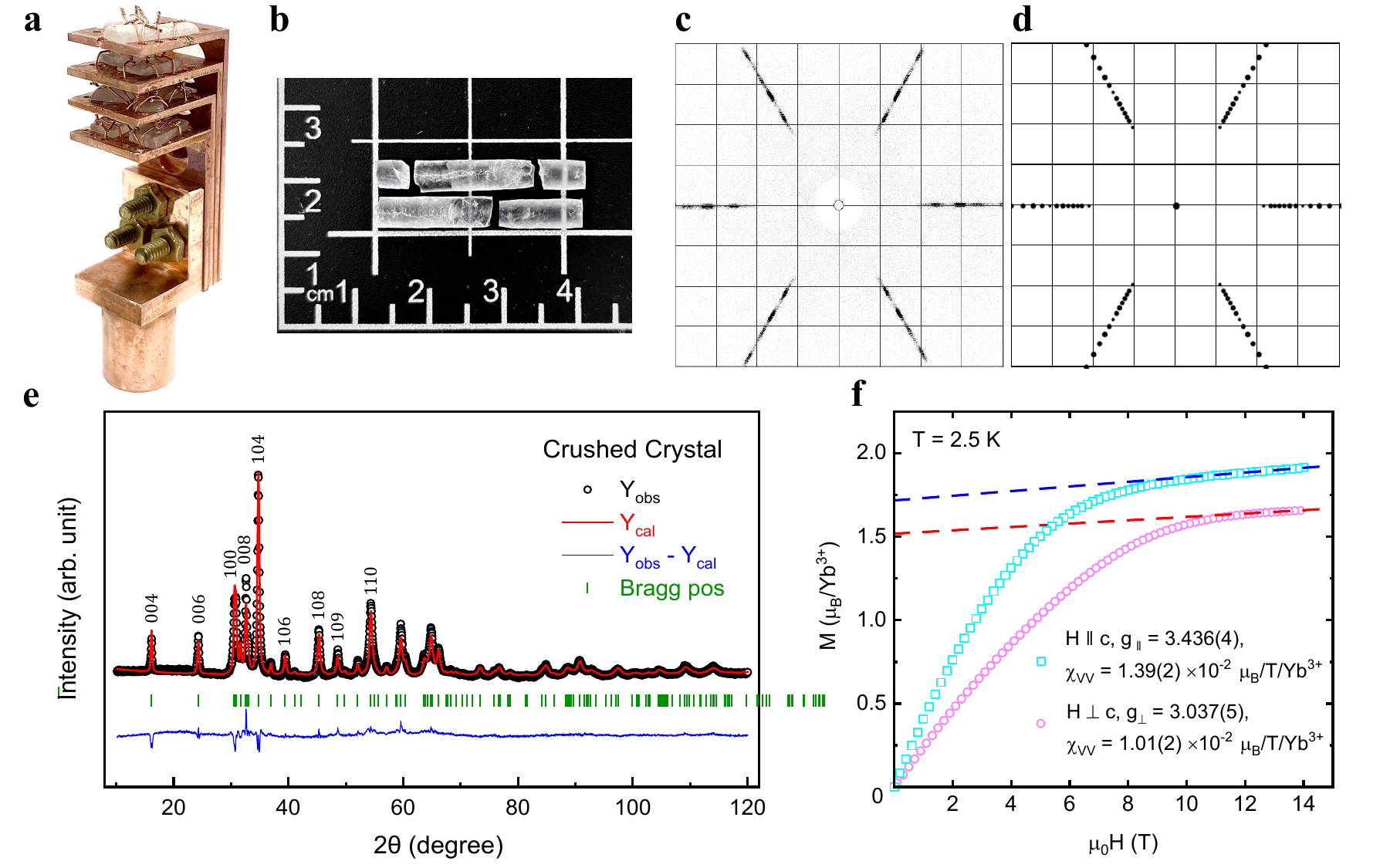}
		\caption{\textbf{Pictures, Laue pattern, powder X-ray diffraction pattern, and isothermal magnetization}. \textbf{a}. A picture of co-aligned single crystals of \YZGO~ on a neutron sample mount. 10 selected high-quality single crystals with a total mass of $\sim$ 1.8 grams were aligned and bound on an oxygen-free Cu sample mount with Cu wire. \textbf{b}. A representative image of the grown single crystals sample of \YZGO, the transparency of the crystal indicates its high quality. \textbf{c \& d}. Experimental and Simulated Laue Back-scattered X-ray diffraction pattern with an incident beam along the \textit{c}-axis. \textbf{e}. Rietveld refinement performed on powder X-ray diffraction (PXRD) pattern of the ground single crystal of \YZGO. \textbf{f}. Field-dependent isothermal magnetization of \YZGO~single crystal collected at 2.5 K. The magnetic field was applied parallel and perpendicular to the crystallographic \textit{c}-axis. The obtained gyromagnetic ratios $g$ and Van-Vleck contributions $\chi_{VV}$ from the fits (red and blue dashed lines) are shown. }
		\label{SFig 1}
	\end{figure*}
	\begin{table}[h]
            \caption{The refinement parameters of the single crystal X-ray diffraction (SCXRD) data collected at 150 K. The atomic coordinates and equivalent isotropic displacement parameters suggest no chemical site mixing in \YZGO. U(eq) is defined as one-third of the trace of the orthogonalized U$^{ij}$ 
            tensor.}\label{SCXRD}
		\begin{tabular} 
            {p{0.8cm}p{0.5cm}p{1.1cm}p{1.1cm}p{1.4cm}p{1.4cm}p{0.8cm}}
		\toprule	
                \multicolumn{7}{l}{\hspace{16mm} YbZn$_2$GaO$_5$ ~ SCXRD~
                ($T = 150\;$ K, Mo$K_{\alpha}$, $\lambda = 0.71073$ \AA) }\\
                \midrule

			\multicolumn{2}{l|}{Space group} &
                \multicolumn{5}{l}{$P6_3mmc$~ (No.194)} \\
                \midrule
                \multicolumn{2}{l|}{\multirow{2}{*}{Cell parameters}} &
			\multicolumn{5}{l}{ $a=b=3.3678(2)$ \AA, ~$c= 21.951(2)$ \AA }\\
                \multicolumn{2}{l|}{} &
                \multicolumn{5}{l}{ $\alpha=\beta=90^{\circ}$, ~$\gamma = 120^{\circ}$ }\\

                \midrule
                \multicolumn{2}{l|}{Fit quality} &
                \multicolumn{5}{l}{$R1 = 0.0779$,~ $wR2 = 0.1895$, ~$GooF = 1.0612$} \\
			
			\midrule
			\textbf{Atom} & \textbf{Site} & \textbf{x(\AA)} & \textbf{y(\AA) } & \textbf{z(\AA) } & \textbf{U(eq)(\AA$^{2}$)} & \textbf{Occ} \\
			\midrule
			Yb1   &  $2a$    & 0.0000     & 0.0000      & 0.5000      & 0.0024    & 1.000       \\
			Ga1   &  $2b$    & 0.0000     & 0.0000      & 0.7500      & 0.0018    & 1.000       \\
			Zn1   &  $4f$    & 0.6667     & 0.3333      & 0.6372(1)   & 0.0018    & 1.000       \\
			O1    &  $4f$    & 0.6667     & 0.3333      & 0.5494(10)  & 0.00018   & 1.000       \\
			O2    &  $2c$    & -0.3333    & -0.6667     & 0.7500      & 0.0032    & 1.000       \\
			O3    &  $4e$    & 0.0000     & 0.0000      & 0.6546(11)  & 0.0020    & 1.000       \\
            \botrule
		\end{tabular}	
	\end{table}
 
	\begin{figure}[htbp]\centering\includegraphics[width=14cm]{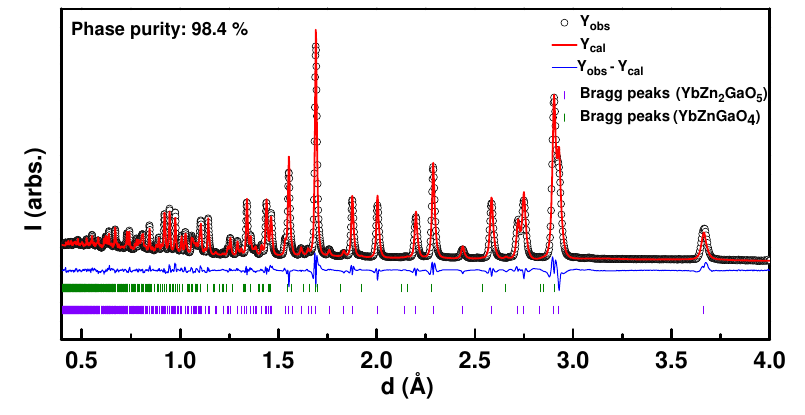}
		\caption{Rietveld refinement was performed on the powder neutron diffraction pattern of \YZGO~collected using the General Materials Diffractometer (GEM) at ISIS Neutron and Muon Source in UK. A small impurity of YbZnGaO$4$, approximately 1.6\%, was observed. Further exploration into the potential site mixing between the Zn and Ga sites was conducted, but no significant improvement to the fit was observed.}
		\label{SFig 2}
	\end{figure}
\begin{table}[h]
\caption{Fitted CEF parameters for \YZGO.}\label{BmnTable}
\begin{tabular}{cc}
\toprule
$B^{m}_{n}$& $\text{Fit values } \, (\mathrm{meV})$ \\
\midrule
$B^{2}_{0}$  &  $-1.008\times10^{0}$  \\
$B^{4}_{0}$  &  $1.459\times10^{-2}$  \\ 
$B^{4}_{3}$  &  $-6.427\times10^{-1}$  \\
$B^{6}_{0}$  &  $5.446\times10^{-4}$  \\
$B^{6}_{3}$  &  $-3.284\times10^{-2}$  \\ 
$B^{6}_{6}$  &  $2.232\times10^{-2}$  \\
\botrule
\end{tabular}
\end{table}
\begin{table}[h]
\caption{Eigenvalues and eigenvectors of the CEF Hamiltonian for \YZGO.}\label{EigenTable}
\begin{tabular}{cc}
\toprule
$\text{Eigenvalues}\, (\mathrm{meV}) $ & $\text{Eigenvectors}$ \\
\midrule
0.00  &   $-0.758 \left| -\frac{7}{2} \right\rangle + 0.402 \left| -\frac{1}{2} \right\rangle + 0.512 \left| \frac{5}{2} \right\rangle $\\
0.00  &  $-0.513 \left| -\frac{5}{2} \right\rangle + 0.402 \left| \frac{1}{2} \right\rangle + 0.758 \left| \frac{7}{2} \right\rangle $\\
38.14  &  $-0.834 \left| -\frac{5}{2} \right\rangle - 0.444 \left| \frac{1}{2} \right\rangle - 0.329 \left| \frac{7}{2} \right\rangle $\\
38.14  &  $-0.329 \left| -\frac{7}{2} \right\rangle + 0.444 \left| -\frac{1}{2} \right\rangle - 0.834 \left| \frac{5}{2} \right\rangle $\\
60.88  &  $\left| -\frac{3}{2} \right\rangle $ \\ 
60.88  &  $\left| \frac{3}{2} \right\rangle $ \\ 
94.00  &  $0.204 \left| -\frac{5}{2} \right\rangle - 0.801 \left| \frac{1}{2} \right\rangle + 0.563 \left| \frac{7}{2} \right\rangle $\\
94.00  &  $-0.563 \left| -\frac{7}{2} \right\rangle - 0.801 \left| -\frac{1}{2} \right\rangle - 0.204 \left| \frac{5}{2} \right\rangle $\\
\botrule
\end{tabular}
\end{table}
	\begin{figure}[htbp]\centering\includegraphics[width=16 cm]{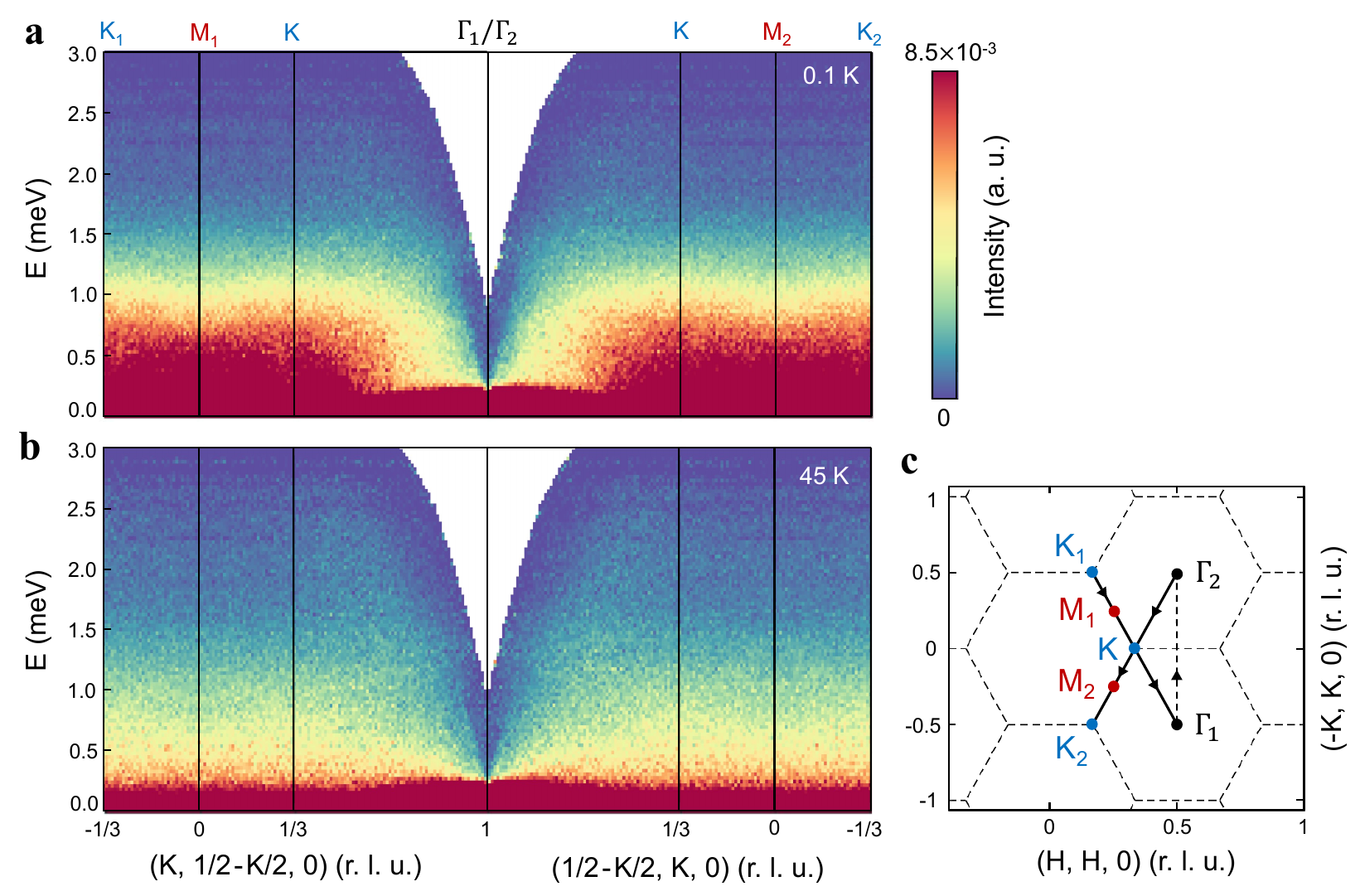}
		\caption{\textbf{S(Q,$\omega$) spectra at 0.1 K and 45 K, with E$_i$ = 3.32 meV}. \textbf{b}. The raw intensity contour plot along high symmetry points measured at 0.1 K, with energy transfer on the y-axis and a reciprocal space path on the x-axis (as depicted in panel c). \textbf{b}. The raw intensity contour plot along the same path was obtained at a higher temperature 45 K. The 45 K spectrum was normalized to 0.1 K data and subtracted as background from the 0.1 K spectrum (as illustrated in Fig. 3a of the paper). \textbf{c}. A schematic of the path used in a and b. The Brillouin zones' boundaries are indicated by dashed lines. The path travels through high symmetry points: $K$$_1$, $M$$_1$, $K$, $\Gamma_1$/$\Gamma_2$, $K$, $M$$_2$, $K$$_2$. }
		\label{SFig 2}
	\end{figure}
        \begin{figure}[htbp]\centering\includegraphics[width=13 cm]{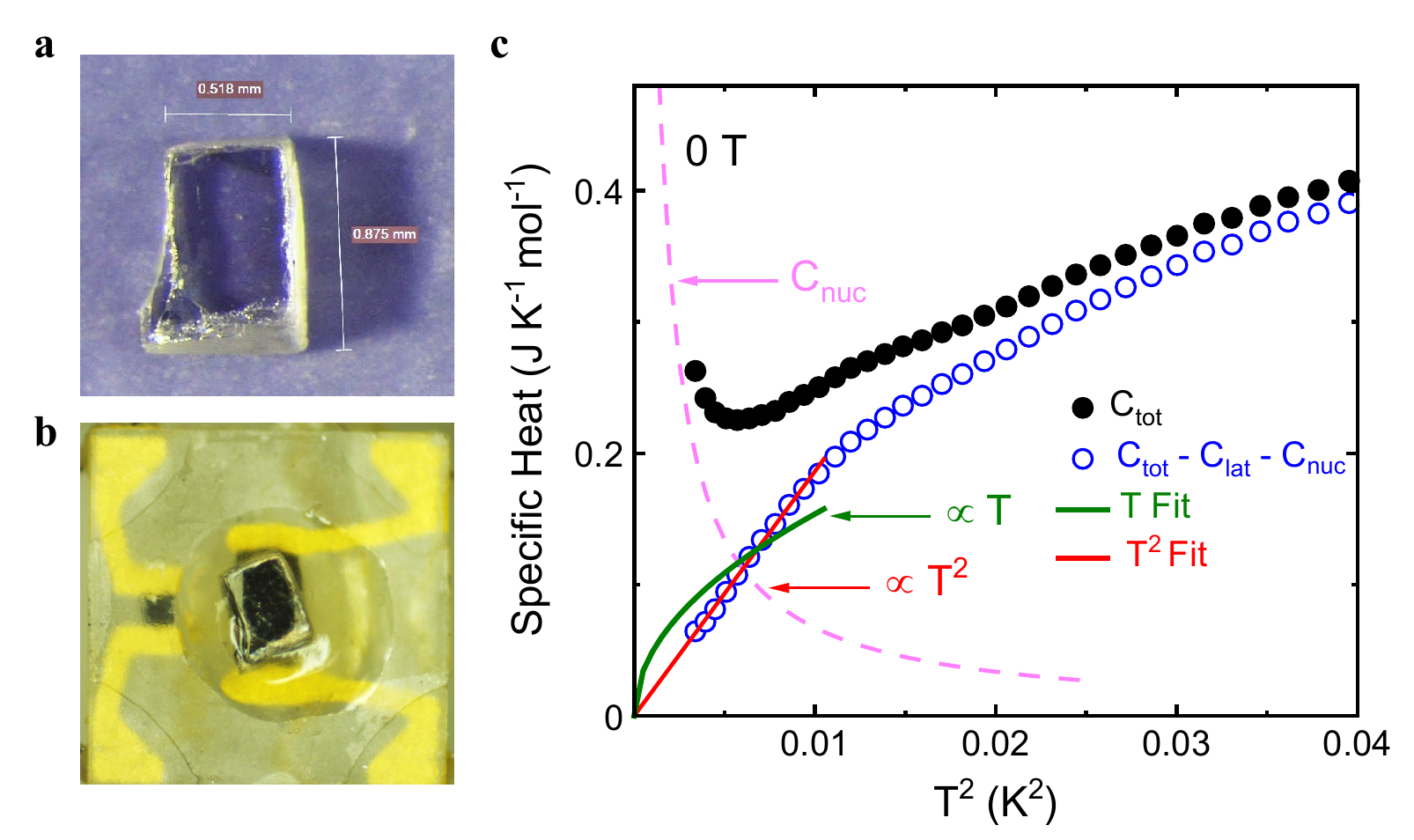}
	\caption{\textbf{Heat capacity data fitted with linear and quadratic T models}. \textbf{a}. Microscopic image of  \YZGO~ single crystal sample used for heat capacity measurement. \textbf{b}. The mounted single crystal sample on a heat capacity platform of the Dilution Refrigerator set up of PPMS. The transparency of the single crystal sample of \YZGO~indicates the crystal quality. \textbf{c}. The filled black circles display the total heat capacity which can be written as $C_{tot} = C_{nuc} + C_{lat} + C_{mag}$. $C_{nuc}$ is expressed as $AT^{-2}$; and $C_{lat}$ is subtracted using the heat capacity of the non-magnetic \LZGO~sample.  The light-magenta dash line indicates the nuclear contribution obtained from the fit. The blue circles show the data after subtracting the nuclear and lattice contribution. The green and red solid lines show the results of subtracted magnetic heat capacity data fitted using a linear ($C_{mag} \sim $T) and a quadratic ($C_{mag} \sim $T$^{2}$) models. The linear T dependence is expected for the spinon Fermi surface quantum spin liquid, while the quadratic behavior for T $\rightarrow$ 0 corresponds to U(1) Dirac quantum spin liquid model.}
		\label{SFig 3}
	\end{figure}

        \begin{figure}[htbp]\centering\includegraphics[width=16.5 cm]{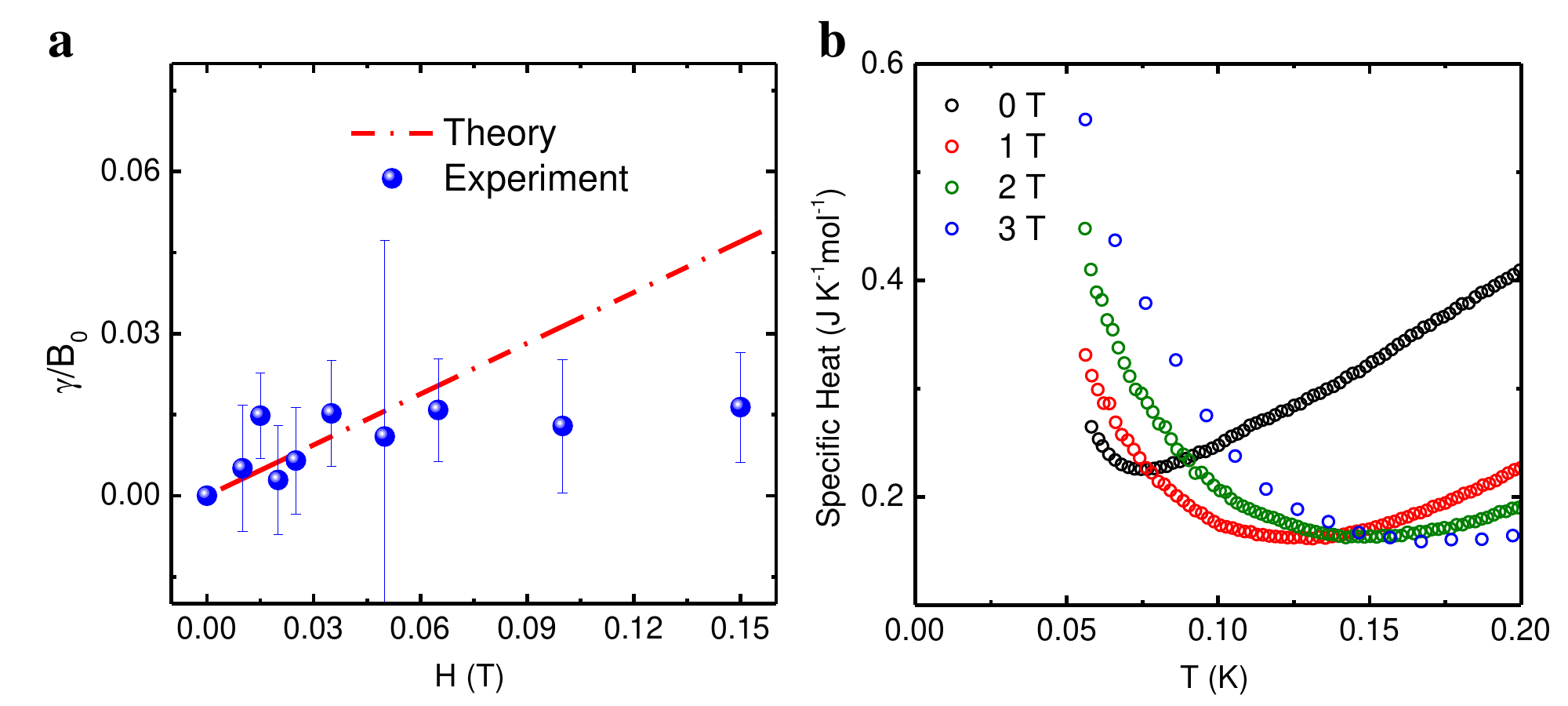}
	\caption{\textbf{Comparison between Dirac QSL theory predictions and experimental results for \YZGO~ under finite applied magnetic fields}. \textbf{a.} Low temperature magnetic specific heat data (C$_\text{mag} = \gamma T + B T^2$) are fitted and the extracted ratios of $\gamma$/$B_0$ (blue points), where $B_0$ is the quadratic term at zero applied field, are obtained from the fitting parameters under various applied magnetic fields. According to the theoretical predictions, as proposed by Ran \textit{et al.} \cite{RanPRL2017}, at the mean-field level; (i) for low temperature limit ($(k_BT \ll \chi J)$), one expects a $C \propto T^2$ law because of the Dirac nodes and (ii) in presence of a finite magnetic field $(k_BT \ll \mu_BH)$, the specific heat goes up linearly with field strength. The calculated theoretical ratio (red dashed line), are obtained from the ratio between eq. \ref{eq:2} and eq. \ref{eq:1} (see note on Fig S5). These results are in good agreement with the experimental fit parameter ratios. As expected, at higher fields outside of Dirac limits, the comparison plot starts deviating from the calculated theoretical values (H). This supports our conclusion that \YZGO ~ realizes Dirac QSL state both at zero field and at finite fields within the Dirac range. \textbf{b.} The total specific heat of \YZGO ~ in the presence of higher magnetic fields of 1, 2, and 3 Tesla.}
		\label{S Fig 5_rebuttal}
	\end{figure}

	\begin{figure}[htbp]\centering\includegraphics[width=13.5 cm]{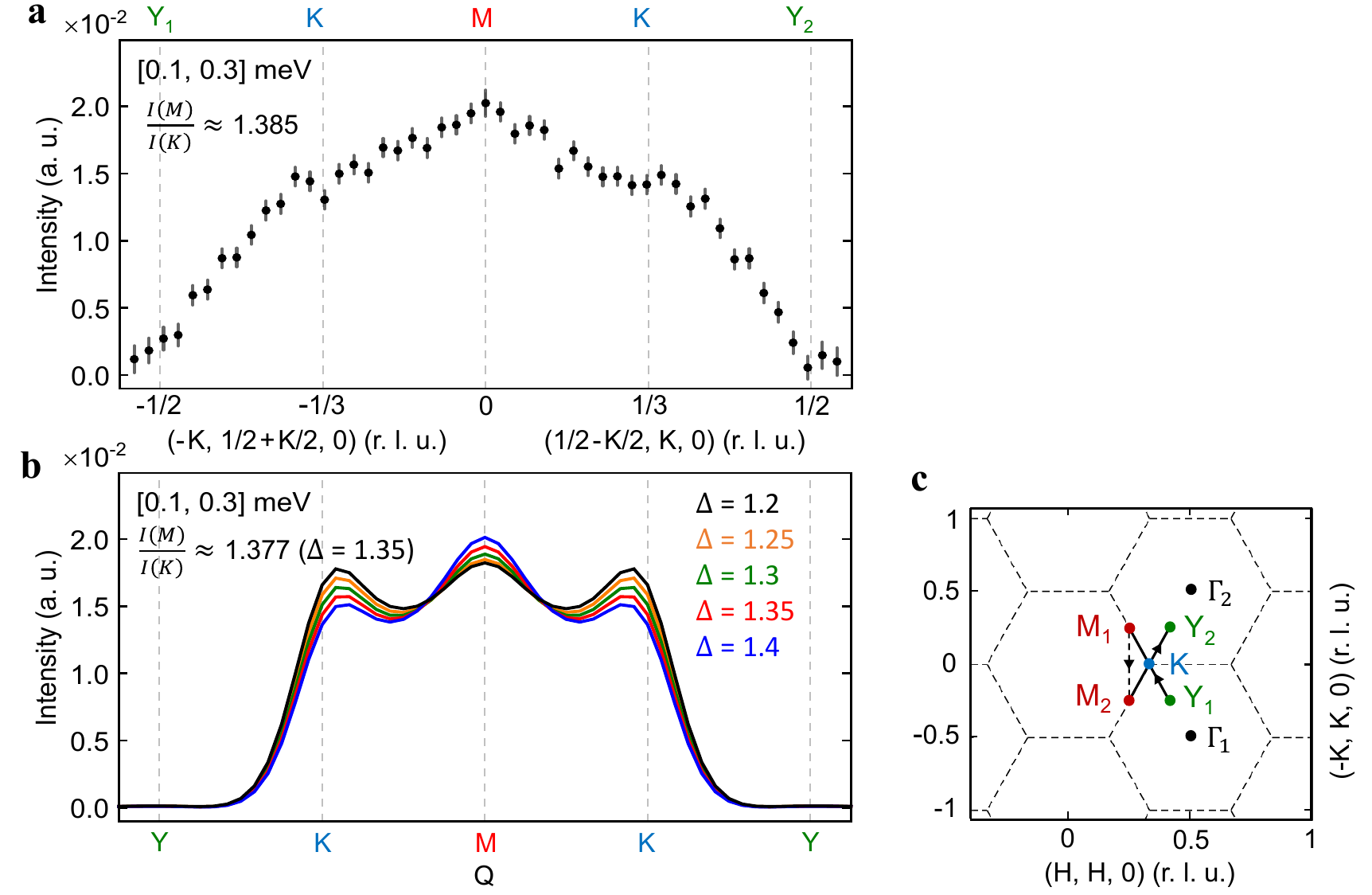}
		\caption{\textbf{Line cut along high-symmetry points and theoretical calculation with different $\Delta$ values}. We present an investigation of the anisotropy effect in the low-energy INS spectrum. \textbf{a}. A line cut traveling through high symmetry points (as depicted in c.): $Y$$_1$, $K$, $M$, $K$, $Y$$_2$. The background-subtracted intensity was integrated between energy intervals [0.1, 0.3] meV and a small q window perpendicular to the high-symmetry path. The ratio of spectral weights between $M$ and $K$ points is around 1.385. \textbf{b}. Theoretical calculations of the spectral weights along the same path using different $\Delta$ values ($\Delta$ = 1.2, 1.25, 1.3, 1.35, 1.4). The ratios between spectral weights at the $M$ and $K$ points for different Delta were calculated, and we find that the ratio of the $\Delta$ = 1.35 case provides the best agreement with our experimental data. \textbf{c}. A schematic for the high symmetry path of the line cut is shown in a and b.}
		\label{SFig 4}
	\end{figure}
	\begin{figure}[htbp]\centering\includegraphics[width=13.5 cm]{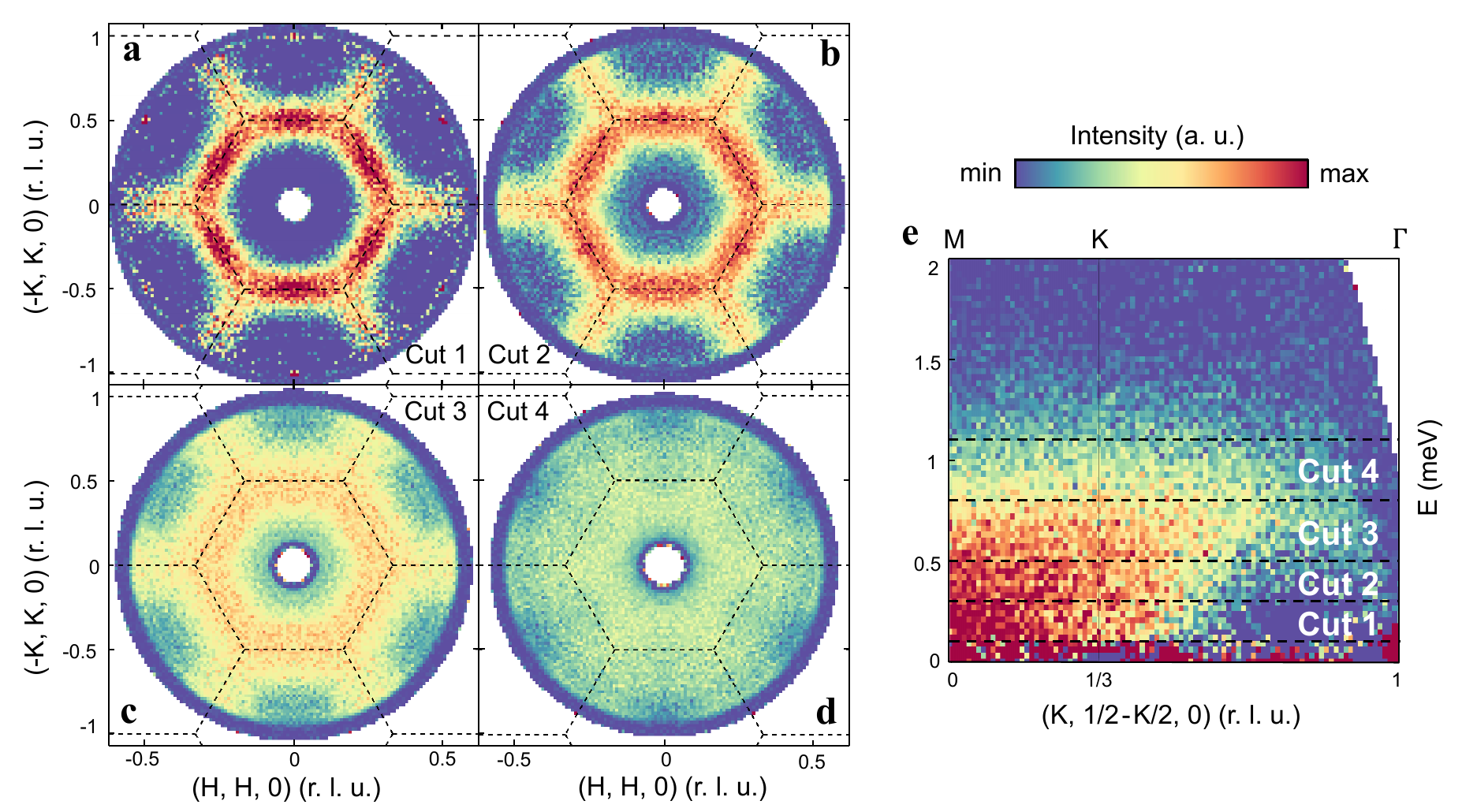}
		\caption{\textbf{Momentum dependence of the magnetic excitation}. \textbf{a-d}. Background-subtracted constant energy slices of the magnetic excitation spectra collected at 0.1 K. The energy integration ranges are [0.1, 0.3],[0.3, 0.5],[0.5, 0.8], and [0.8, 1.1] meV. The spectra were folded along high symmetry directions to increase statistics and the folded data was duplicated and recombined to restore full coverage for presentation, as explained in the Methods section. The spectral weights are expressed by the same color map scale for direct comparison. \textbf{e}. S(q,$\omega$) plot along high symmetry path $M$-$K$-$\Gamma$. The dash lines indicate the energy integration intervals of cuts 1-4 for better visualization.}
		\label{SFig 5}
	\end{figure}
\clearpage

\textbf{Note on Fig S5: Comparison between Dirac QSL theory prediction and experimental results obtained for \YZGO}

\setlength{\parskip}{1em}

Based on the theoretical predictions for Dirac QSL state, reported by Ran \textit{et al.} \cite{RanPRL2017}, at the mean-field level the calculated specific heat can be expressed as follows:

\text{(i)} At low temperatures ($k_BT \ll \chi J$) and zero applied magnetic field, one expects a $C \propto T^2$ law due to formation of Dirac nodes. In this case $k_B$, $T$, $\chi$ and $J$ are Boltzmann constant, temperature, magnitude of the self-consistent mean-field parameter, and the coupling constant, respectively. In this limit,  the calculated specific heat is:

\begin{equation}
\quad \frac{C}{T^2} = \frac{72\zeta(3)k^3_BA}{\left(2\pi\hbar\right)^2} = 1.1 \chi^{-2} \times 10^{-3} \text{ Joule/molK}^3 \quad \text{(for } k_BT \ll \chi J \text{)}
\label{eq:1}
\end{equation}

\text{(ii)} When subject to finite magnetic field (H), the spinon create a Fermi pocket, with the radius is directly proportional to the magnetic field strength. Hence, at low temperature limit and under finite applied magnetic field ($k_BT \ll \mu_BH \ll \chi J$), one can expect $C \propto T$ law. In this case $\chi$, $J$, $A$, $v_F$, $H$, $\mu_B$ and $k_B$ are the magnitude of the self-consistent mean-field parameter, the coupling constant, the area of the two-dimensional system, the Fermi velocity, external magnetic field, Bohr magneton, and the Boltzmann constant, respectively. In this limit, the calculated specific heat is:

\begin{equation}
\quad \frac{C}{T} = \frac{8\pi^3 k^2_BA\mu_BH}{3\left(2\pi\hbar v_F\right)^2} = 0.23 \chi^{-2} H \times 10^{-3} \text{ Joule/molK}^2 \quad \text{(for } k_BT \ll \mu_BH \text{)}
\label{eq:2}
\end{equation}

The low-temperature specific heat data of \YZGO ~ are collected under several magnetic fields, starting from 0.01 T to 3 T. To find out the fitting parameters in the presence of finite magnetic fields, the magnetic specific heat data (C$_{mag}$) is fitted using $\gamma T$ + $B T^2$ . The extracted parameters ($\gamma$ and $B$) are shown in Fig. 2(f) of the main manuscript.  

The theoretical ratio between equations \ref{eq:2} and \ref{eq:1} gives  $\left(\frac{0.23 \chi^{-2} H \times 10^{-3}}{1.1 \chi^{-2} \times 10^{-3}}\right)$. This ratio can be compared to the experimental fit parameter values ($\gamma$/$B_0$) obtained from the magnetic specific heat data. The $B_0$ is the quadratic fit parameter ($B_0 T^2$) obtained from fitting the magnetic specific heat data of \YZGO ~ at zero applied magnetic field. 

According to the proposed theory by  Ran \textit{et al.}\cite{RanPRL2017}, the Dirac QSL behavior is expected at low temperatures. Our experimental specific heat data is collected down to 0.06 K ($\simeq$ 0.04 Tesla) which is the lowest temperature limit in our Dilution Refrigerator (DR) experimental setup. However, to verify the Dirac QSL theory ($k_BT \ll \mu_BH \ll \chi J$), we collected low-field specific heat data starting from 0.01 T, 0.015 T, 0.025 T and 0.05 T, with small temperature increments. The extrapolated the datasets down to zero are in good agreement with the calculated theoretical values, as indicated by a linear increase in (C$_{mag}$/$T$)  with increasing applied magnetic field strength (see Fig. 2e of the main manuscript).

Furthermore, as shown in Fig. S5a, the calculated theoretical ratio (plotted in red) is in good agreement with the experimental ratio ($\gamma$/B$_0$, represented by the blue data with error bars) in the low field limit ($k_BT \ll \mu_BH$). However, with increasing applied magnetic field strength (H), the experimental ratio begins to deviate from the theoretical ratio. This observation supports our conclusion that the compound \YZGO is a promising candidate for a Dirac QSL.

\bibliographystyle{apsrev4-2}
\bibliography{Yb1215_SM.bib}